\documentclass[11pt,twoside]{article}

\usepackage{asp2006}
\usepackage{epsf}

\usepackage{lscape}
\usepackage{graphicx}

\begin{document}
\title{A new grid of evolutionary synthesis models}
\author{M. Moll\'{a},$^{1}$ M.L. Garc\'{\i}a-Vargas,$^{2}$
 Alessandro Bressan,$^{3}$  and Pedro G\'{o}mez-Alvarez $^{4}$}

\affil{$^{1}$ CIEMAT, Avda. Complutense-22, 28040 Madrid, (Spain)}
\affil{$^{2}$ FRACTAL, Castillo de Belmonte-1,  28232 Las Rozas de Madrid, (Spain)}
\affil{$^{3}$ INAF Osservatorio Astronomico di Padova, 
Viccolo dell' Osservatorio-5,  35122 Padova (Italy)}
\affil{$^{4}$ ESAC, Villafranca del Castillo, P.O. Box 50727, 
28080 Madrid (Spain)}
 
\begin{abstract}
We present new evolutionary synthesis models for Single Stellar
Populations covering a wide range in age and metallicity. The most
important difference with existing models is the use of NLTE
atmosphere models for the hot stars (O, B, WR, post-AGB stars, and
planetary nebulae) that have an important impact in the stellar
cluster's ionizing spectra.  
\end{abstract}

\section{Synthesis Code Description}  

We have used the synthesis code by \cite{gmb98}, updated by
\cite{mgv00} and newly revised now.The basic grid is composed by
Single Stellar Populations (SSP) for five different IMF's. The first
one is a \cite{sal55} power law with masses between 0.85 and 120 M$_{\odot}$.  
The others IMFs are \cite{sal55}, \cite{fpp90}, \cite{kro02} and 
\cite{cha03} functions, with masses between 0.15 and 100 M$_{\odot}$.  

The isochrones are those from \cite{bgs98} for 6 different metallicities: 
Z $=$ 0.0004, 0.001, 0.004, 0.008, 0.02 and 0.05.  
The age coverage is from $log{t}=$ 5.00 to 10.30 with a variable time
resolution which is $\Delta(log{t})=0.01$ in the youngest stellar
ages. The WC and WN stars are identified in the isochrones according
to their surface abundances.

The atmosphere models are from \cite{lcb97} with an excellent coverage
in effective temperature, gravity and metallicities, for stars with
Teff$ \leq 25000$ K.  For O, B and WR we have taken the NLTE blanketed
models by \cite{snc02} at Z$=$0.001, 0.004, 0.008, 0.02 and
0.04. There are 110 for O-B stars, with 25000 K $ < Teff \leq 51500$ K
and $2.95 \leq \log{g} \leq 4.00$, and 120 for WR stars (60 WN $+$ 60
WC), with 30000 K $\leq T^{*} \leq 120000$ K and $1.3R_{\odot}\leq
R^{*}\leq 20.3 R_{\odot}$ for WN, and with $ 40000 K \leq T^{*} \leq
140000$ K and $ 0.8R_{\odot}\leq R^{*}\leq 9.3 R_{\odot}$ for
WC. T$^{*}$ and R$^{*}$ are the temperature and the radius at a
Roseland optical depth of 10.

To assign a model to each WR star, we use the relationships among
opacity, mass loss and wind velocity: $d\tau=-\kappa(r)\rho(r)dr$,
where $\kappa(r)=-0.2(1+X_{\rm S})$ and $X_{\rm S}$, the H surface
abundance, is taken as 0.2 for WN and 0 for WC.
The mass loss is: $\frac{dM}{dt}=4\pi
r^{2}\rho(r)v(r)dr$ with $v(r)=v_{\infty}(1-R_{S}/r)^{\beta}$ taking
$\beta=1$.  Integrating these equations we find R$^{*}$ and then we
select the closer atmosphere model.

For post-AGB and PN with Teff from 50000 to 220000 K we take the NLTE
models by \cite{rau03}. For higher temperatures we use black bodies.

\section{Results}
\begin{figure*}[!ht]
\begin{center}
\includegraphics[width=0.45\textwidth,angle=90]{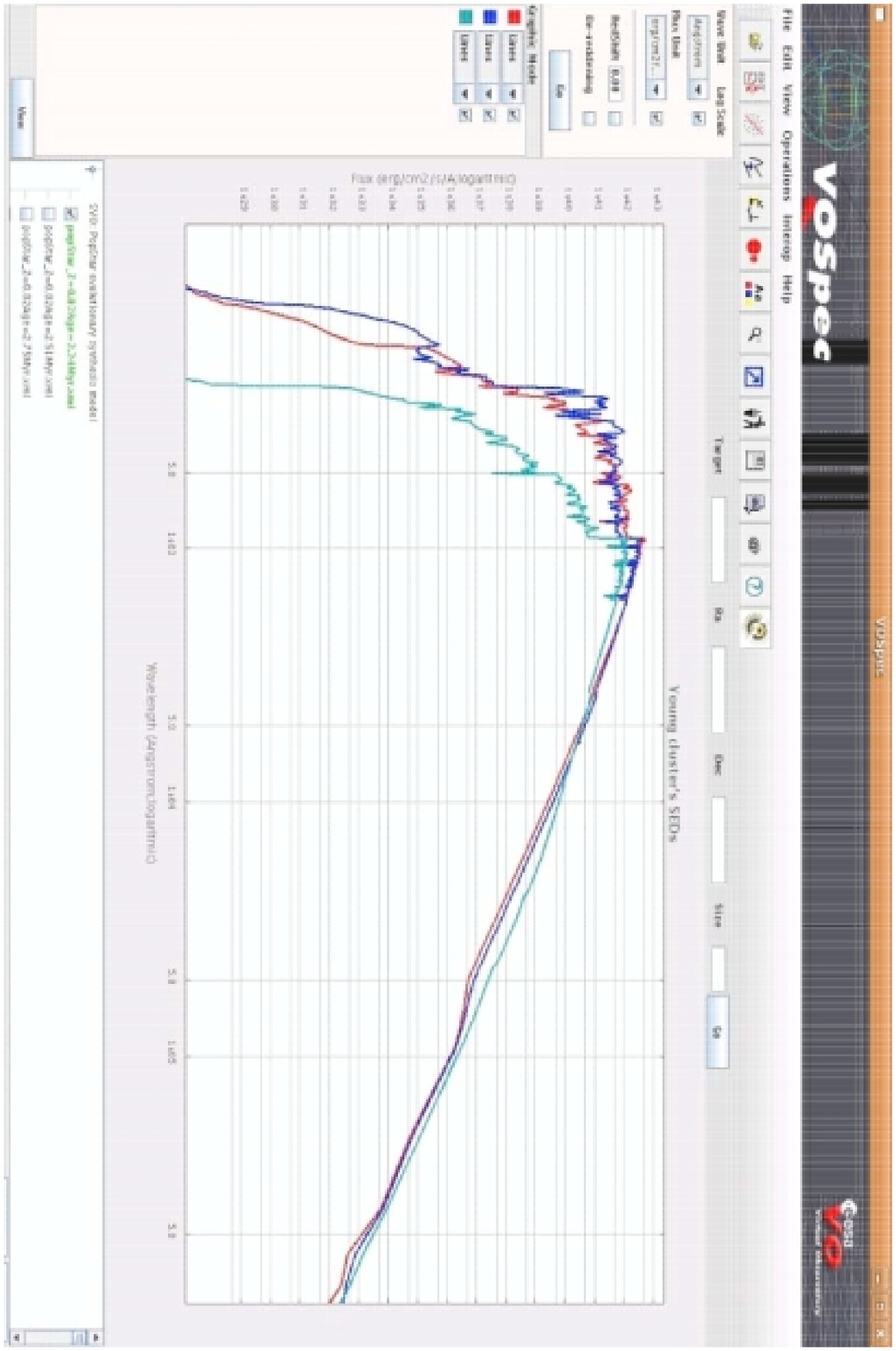}
\caption{SEDs for some young clusters as they are seen in VOspec.}
\end{center}
\end{figure*}

The resulting SEDs and HR diagrams \footnote{Models 
in http://esavo.esa.int/vospec/ and
http://www.fractal-es.com/SEDmod.htm} are available in the VO.  The
use of NLTE blanketed models produce less hard ionizing photons than
old models \cite[e.g.][]{gbd95} which explains in antural way the
emission line ratios in low excitation high metallicity H{\sc ii}
regions. Previous work needed a steeper IMF or mass segregation in
small cluster, in disagreement with evidences from HST of the
existence of very massive stars even in small clusters.

\end{document}